# Data-Driven Innovation: What Is It?

**Jianxi Luo**

Data-Driven Innovation Lab

Singapore University of Technology and Design

Email: luo@sutd.edu.sg

**Abstract**

The future of innovation processes is anticipated to be more data-driven and empowered by the ubiquitous digitalization, increasing data accessibility and rapid advances in machine learning, artificial intelligence, and computing technologies. While the data-driven innovation (DDI) paradigm is emerging, it has yet been formally defined and theorized and often confused with several other data-related phenomena. This paper defines and crystalizes "data-driven innovation" as a formal innovation process paradigm, dissects its value creation, and distinguishes it from data-driven optimization (DDO), data-based innovation (DBI), and the traditional innovation processes that purely rely on human intelligence. With real-world examples and theoretical framing, I elucidate what DDI entails and how it addresses uncertainty and enhance creativity in the innovation process and present a process-based taxonomy of different data-driven innovation approaches. On this basis, I recommend the strategies and actions for innovators, companies, R&D organizations, and governments to enact data-driven innovation.

# 1. Data-Driven Innovation as a Process Paradigm

Innovation addresses evolving societal needs and drives business and economic growth. However, innovation is never easy or assured, because of the innately uncertain nature of the innovation process. The aim of innovation is to create new products, services and systems, new demands, and new markets. In turn, the creativeness and newness that define innovation also result in uncertainty in the process of pursuing innovation. It is often ambiguous what to innovate and how to innovate. Uncertainty surrounds the innovation process and challenges all innovators and the companies and governments that aspire to innovate.

Meanwhile, the digitalization of our work and life have been generating growing data about technologies, processes, users, markets, etc., daily, hourly, and second-by-second. If we can properly mine, analyze and make sense of such data, particularly the naturally and often passively generated digital footprint data, we will be able to make innovators more inspired and innovation decisions more informed, and thus reduce the uncertainty and increase the creativity in the innovation process. This is the "Data-Driven Innovation (DDI)" process.

Herein, I define data-driven innovation as the process of innovating that draws information and inspiration from big data (of users, innovators, stakeholders, science and technologies, processes, environments, etc.). The data-driven information and inspiration can reduce uncertainty and enhance creativity in the innovation process (see how it works in Section 3). Data science, machine learning, artificial intelligence (AI), and computing technologies further empower data-driven innovation.

The term "data-driven innovation" has appeared in both lay press and a small number of reports and papers in the past few years but was given ambiguous or different meanings [1,2]. The concept is often confused with the innovative products and services whose core features and value creation for users are based on data (which we will call "data-based innovation") [3,4,5], or with the operations which mine and analyze data to optimize efficiency and accuracy in delivering pre-defined objectives (which we will call "data-driven optimization") [6,7,8]. Such ambiguity might limit or misguide the actions and efforts aimed to make the innovation processes more data-driven, informed and inspired.

Herein, this article aims to crystally define data-driven innovation (DDI) as a process and distinguish it from other data-related paradigms. I will first start with a few real-world examples of DDI and theoretical framing of what it entails and how it creates values. On this basis, I present a taxonomy of different data-driven approaches according to the values they create for different exploratory and creative actions in the innovation process. Finally, I discuss relevant technological, organizational and policy considerations to embrace the future of innovation processes that are data-driven and AI-empowered.

## 2. What Is Data-Driven Innovation: Some Real-World Examples

One example of data-driven innovation is when Intuit created Quickbooks, an accounting management software for small businesses. The idea was conceived when the team analyzed the digital footprint data of users for their earlier product Quicken, a personal financial management software, and found many customers used Quicken in workplaces for business accounting. Then the team developed a new product specifically for small business accounting, which was the Quickbooks [9]. In this case, the inspiration for innovation was obtained from rather simple user digital footprint analytics. In fact, DDI can be further empowered by advanced big data computation, machine learning and artificial intelligence technologies.

For instance, IBM conducted a large-scale semantic analysis of millions of public text documents about diverse materials to identify three natural materials which have never been used in a battery and are unknown by battery engineers and combined them in a new battery design that outperforms state-of-the-art lithium-ion batteries [10]. In another example, Atomwise, a San Francisco-based drug discovery startup, trained neural networks on large-scale prior experimental data to predict the performance of new drugs to accelerate drug innovation [11]. Since the outbreak of COVID-19, researchers around the world have been racing to develop medical cures and vaccines. Many have taken a data-driven approach to mine large databases of compounds, peptides, and epitopes to train machine learning models to discover, generate and evaluate vaccines and candidate cure medicines [12].

Enormous data that can be mined to aid innovators are publicly available and accessible everywhere in our digitalized life. For instance, e-commerce websites host massive publicly accessible data about products and their design specifications, users, and their opinions about products. Researchers have mined data from e-commerce websites to train neural networks that can learn the latent needs of consumers, based on their comments on consumer products and translate such into technical design requirements to inform product design [13]. In fact, mining and analyzing consumers' behavior and opinion data from e-commerce sites, social media, and other online or digital spaces to inform new product development has been a common practice and well-studied in the marketing literature.

In the meantime, the digital footprints of innovators and innovation activities may also inform and inspire innovators and companies for innovation. For example, the public patent database contains enormous unstructured multimodal descriptions of prior technologies across domains and digital footprints of inventors and companies regarding their innovation behaviors and performances. There have been efforts to develop patent data-driven expert systems to inform companies of latent innovation opportunities specifically for them [14] and retrieve and prompt design stimuli from the patent database to inspire innovators for generating new design ideas [15], as well as to train neural networks with patent data to predictively evaluate the value of new inventions [16].

The US Chamber of Commerce Foundation [2] highlighted four types of open big data that can be explored to generate value across sectors. They include user-generated data (often online), industrial and sensor data (from GPS, mobiles, machines/equipment, and public infrastructure), business or enterprise data (such as inventories and transactions), and public data (curated or generated by government agencies, universities, or non-profit organizations, such as patents, papers, reports). Such open and public data sources, as well as the proprietary data of companies and government agencies, may enable enormous opportunities for data-driven innovation.

## 3. How Does DDI Create Value: A Theoretical Explanation

*3.1 Dissecting the Innovation Process*

These aforenoted examples demonstrate various ways or manners in which data-driven approaches can inspire, inform, and augment a variety of exploratory or creative but uncertain actions in the innovation process. Here we further synthesize them to present a taxonomy of different data-driven approaches according to the different values they create for the complementary actions in typical innovation processes. To aid the synthesis, we focus on four core innovation process actions that shape the opportunity space and the design space essential for innovation (Figure 1).

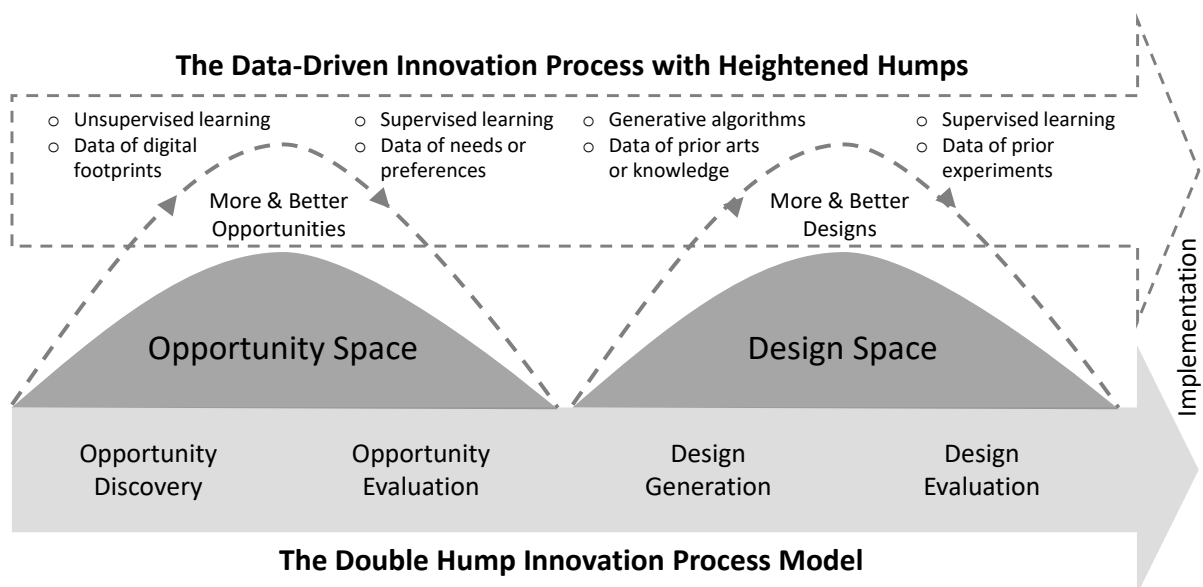

**Figure 1.** The Double Hump Model Explains Value Creation of the Data-Driven Innovation Process.

The opportunity space is the set of alternative opportunities where the innovators want to and can create value via designing new solutions, products, services, or systems to address human or societal needs. The opportunity space exits latently but needs to be discovered via the exploration of latent human needs that are unmet or new and better uses of sciences and technologies. Then the discovered alternative opportunities need be evaluated and compared to specify the opportunity to address and define it with clear requirements to guide design generation. The divergent opportunity discovery action, followed by the convergent opportunity evaluation action, gives rise to the hump representing the opportunity space in Figure 1.

The design space is the set of design alternatives that the innovators come up with to address the specified opportunity. The design space does not exist previously and needs to be created via the generation of many, diverse designs. Here the "designs" refer to the new and useful artefacts with societal and economic values, instead of pure art designs for self-expression. Then the generated new design alternatives need to be evaluated and compared to select the design for use implementation. The divergent design generation action, followed by the convergent design evaluation action, gives rise to the hump representing the design space in Figure 1.

Hereafter, I call the innovation process representation in Figure 1 the *double hump model*. The two humps represent the opportunity and design spaces respectively. It is noteworthy that, real-world innovation processes involve additional actions, such as implementation and entrepreneurship [17], which might also condition innovation success. The *double hump* model focuses on the core actions for innovation. Moreover, the two-hump four-action process also involves feedback and iterations not shown in Figure 1. For instance, one might perceive additional opportunities during opportunity evaluation, design generation and evaluation. New design ideas can be conceived with the feedback and learning obtained during design evaluation.

Higher opportunity and design space humps indicate greater upper bounds of the creativity and value creation of the innovation process. Creativity is the ability to create novel and useful artefacts [17, 18, 19]. Innovation relies on creativity [20, 21, 22]. A more creative innovation process is expected to be able to generate more novel and useful designs. While innovators often, due to resource constraints, only acuate a small number of designs to address a small number of opportunities, discovering a more diversified scope of opportunities and generating a greater variety of design alternatives at the first place may give rise to the variance of opportunities and designs and thus increase the novelty and usefulness of the best opportunity and the best design in the end.

*3.2 How Data-Driven Approaches Can Create Value for Innovation*

Thus, to heighten the humps, greater divergence in opportunity discovery and greater variety in design generation is beneficial. In result, the heightened humps (representing the expanded opportunity and design spaces) would further require more efficient opportunity and design evaluation to converge towards the final opportunity to design for and the final design to implement. The *double hump* model, by highlighting the specific divergent and convergent actions in the innovation process, provides a taxonomy (Table 1) to synthesize the different data sources and data-driven methods (see examples in Section 2) according to the respective actions they can augment.

**Table 1.** Data-Driven vs. Human-Social Approaches to Innovation Process Actions

| Innovation Process Actions | Data-Driven Approaches | Human-Social Approaches |
|---|---|---|
| Opportunity Discovery | - Unsupervised learning<br>- Digital footprint data about users, innovators, stakeholders, etc.<br>- Enlarge opportunity space | Interviews, surveys, observations of people to discover problems, frustrations, and challenges as innovation opportunities |
| Opportunity Evaluation | - Supervised learning<br>- Data of user needs or preferences<br>- Accelerate opportunity evaluation | Intuitive human expertise is required to specify design requirements based on user needs or preferences. |
| Design Generation | - Generative algorithms<br>- Databases of prior designs or knowledge<br>- Enlarge design space | Individual design ideation, gut feeling and social processes such as brainstorming and crowdsourcing to generate new design ideas |
| Design Evaluation | - Supervised learning<br>- Data of prior experiments<br>- Accelerate design evaluation | Experiments with users engaged to evaluate new designs |

For *opportunity discovery*, the traditional approaches involve interviewing, surveying, or observing people to develop empathy and understand their needs, problems, frustrations, and challenges. By taking a data-driven approach, innovators may mine and analyze the users' digital footprints of using or commenting on existing products or services to gain insights on the unmet needs (e.g., Quicken [9]). Innovators can also explore the digital footprints of their own in the innovation process to gain insights on the unexplored applications of their technologies as innovation opportunities (e.g., InnoGPS [14]). Opportunity discovery is an exploratory action by nature and thus best supported by unsupervised machine learning techniques.

For *opportunity evaluation*, innovators often rely on human expertise to translate the insights about people and needs gained from the discovery phase into specific design requirements that can direct design generation. In contrast, a data-driven approach can better cope with the opportunity uncertainty especially when the latent opportunity space is large and complex, by making innovators more informed when they evaluate and select opportunities for further pursuits [23]. For instance, innovators may utilize the digital footprint data of user preferences together with the design specifications of the products they purchased, used, or liked (e.g., from e-commerce sites) to train artificial neural networks that relate the former to the latter. Then such trained artificial neural networks can be used to automatically translate newly discovered user needs to specific design requirements [13]. Such actions are translational by nature and thus best supported by supervised machine learning techniques.

For *design generation*, traditional processes rely on human expertise, gut feeling, or social processes such as brainstorming and crowdsourcing to tap on crowd human intelligence. By taking a data-driven approach, innovators may obtain inspirations from data repositories of prior designs and inspiration sources, e.g., asknature.org, moreinspiration.com and patent database that store the knowledge and concepts from millions of inventors, to stimulate creative ideation [15,24]. Further, computer algorithms can automatically generate new design concepts by mutating, combining, and recombining prior concepts from large knowledge databases [25]. For instance, genetic programming algorithms may start from a small set of initial designs to generate alternatives that better satisfy a pre-defined fitness function [26]. With the rapid advances of deep learning, the neural network-based Generative Adversarial Networks (GAN) and Generative Pre-trained Transformers (GPT) have shown the ability to efficiently generate many and more novel designs that deviate from the training data [27,28].

For *design evaluation*, traditionally companies and innovation teams develop minimum viable products and engage external lead users for testing and feedbacks. Such processes are expensive, time-consuming, and serendipitous. By taking a data-driven approach, innovators can automatically evaluate and validate a very large quantity of diverse design concepts to accelerate the design evaluation, validation, and selection process. For instance, innovators can automatically evaluate and filter many new design concepts with a pre-trained common-sense knowledge base [29]. One can also train deep neural networks with the data of prior experiments or successful/failed designs in the same context for automatically and predictively evaluating the performances and value of next design concepts (e.g., Atomwise [11]; [16]).

In summary, unsupervised learning of digital footprint data may help explore many more alternative opportunities and discover a bigger opportunity space than what humans could perceive on their own. The expanded opportunity space may cover more novel and more valuable candidate opportunities to choose and pursue and thus benefit creativity. Innovators drawing inspirations from large knowledge databases (e.g., patents, papers) and using data-driven generative algorithms may generate many design alternatives and create a bigger design space than what humans alone could conjecture up. The expanded design space may contain more novel and more useful candidate designs to choose and implement and thus benefit creativity. Meanwhile, data-trained models can automate the evaluation of many opportunities [23] and many design concepts [29] from the enlarged opportunity and design spaces, accelerate the convergent search and ensure the identification of the best innovation opportunity to design for and the best design for implementation.

Therefore, both the divergence-oriented actions (opportunity discovery and design generation) and the convergence-oriented actions (opportunity evaluation and design evaluation) can be augmented by different suitable data-driven approaches (See the taxonomy in Table 1) to achieve greater creativity of the innovation process.

# 4. Distinguishing Data-Driven Innovation (DDI) from Data-Based Innovation (DBI) and Data-Driven Optimization (DDO)

To define data-driven innovation as a formal process paradigm, we need to distinguish it from other seemly related paradigms or phenomena. Data-driven innovation is the process of innovating that is data-driven to make innovators and their activities and decisions more informed, inspired, and more creative – whether by applying data science, machine learning or AI to large-scale data of users, innovators or more stakeholders. It contrasts with the traditional innovation process that relies on human expertise (creative genius), social activities (brainstorming, crowdsourcing, open innovation), and serendipity (see Table 1).

The data-driven innovation (DDI) paradigm should not be confused with data-based innovation (DBI), which refers to an innovation process's creative output artefacts that are data-based, such as Tik Tok, Coursera, Google Maps, Siri, and those related to the data-based features of IoT (Internet-of-Things) and other physical devices, equipment, machines, or infrastructure [3,4]. DBI benefits users with additional data-based utilities, such as analytics, search, recommendation, and Q&A, in the use process. DBI may result from an intuitive and human-social innovation process, or a data-driven innovation process. DBI can benefit DDI when DBI generates digital footprint data of users during their use process to feed and fuel a further data-driven innovation process, although DDI may also utilize other types or sources of data.

DDI is also distinct from data-driven optimization (DDO) of operations, examples being using demand data and forecasts to inform procurement decisions for a supply chain operation, analyzing real-time traffic data to recommend nearest drivers in a ride hailing service, analyzing users' search behaviors to target online advertising, and analyzing stock market data to inform trading decisions. DDO requires well-defined objective functions, decision variables, and constraints, and often uses real-time data for specific variables to optimize the predefined objective functions, for greater operational efficiency, service delivery quality, and financial returns. In contrast, DDI is open-ended and aims to augment creativity to invent something new. DDI processes may discover new objectives and/or define new decision variables and create radically new values in the undefined whitespace.

While DDI, DDO, and DBI are all enabled by digitalization and computing, machine learning, and AI advances, they create different values for different agents in different types of actions or processes (Table 2). DDI addresses the uncertainty challenge facing innovators in creative processes and enhances the creativity for innovation, as explained in the foregoing section. DDO optimizes the pre-defined decisions of the operators in the operational process toward pre-defined operational objectives. DBI increases the utility of products and services for users in the use process.

**Table 2.** Differentiating the Data-X Paradigms

|  | **Data-Driven Innovation (DDI)** | **Data-Based Innovation (DBI)** | **Data-Driven Optimization (DDO)** |
|---|---|---|---|
| **Process** | *Creative Process* | *Product/Service Use Process* | *Operational Process* |
| **Agent** | *Innovator* | *User* | *Operator* |
| **Value** | *Creativity* | *Utility* | *Optimization* |

## 5. Evolution of Innovation Process Paradigms: From the Past to the Future

The innovation processes have evolved constantly over human history. In the 19th century, innovations came from individual creative geniuses, such as Thomas Edison, Alexander Bell, and Nikola Tesla. In much of the 20th century, well-organized processes in formal R&D labs and centers in large companies such as IBM and AT&T championed innovation. Since the 1970s, innovation ecosystems like Silicon Valley, where frequent and inexpensive knowledge flows across organizational boundaries, have emerged as the powerhouse of innovation.

Today, companies seek innovative opportunities and ideas via various proactive open innovation programs. They collaborate with university-based researchers, organize corporate venture labs jointly with external venture capitalists and entrepreneurs, and run open innovation contests, idea crowdsourcing campaigns, and hackathons to tap into collective and crowd intelligence. To date, the innovation processes have mainly relied on human intelligence (of either individuals, human crowds, organizations, or ecosystems), with high serendipity.

As the society continually creates new knowledge and invent new technologies, we accumulate an expanding space of prior arts and knowledge, which can further benefit future efforts to innovate if they can be efficiently searched, retrieved, learned, and synthesized. Otherwise, the expanding space of precedents may instead appear as knowledge burdens on future innovators, who would be required to learn and know a great deal more precedents before they can design new and more useful products beyond the prior ones.

The growing knowledge burdens [30] will make the innovation process that solely relies on human intelligence less effective in the future. In contrast, data computation and rapidly advancing AI capabilities [31] can quickly and tirelessly learn an enormous knowledge from large unstructured datasets and transform or synthesize the learned knowledge to generate design concepts of new products and services and automatically evaluate them. Artificial intelligence fueled with big data is driving us into the new "decade of technology intelligence [32]".

Therefore, the key enabler for DDI down the road will be the AI for creative tasks, namely, "Creative Artificial Intelligence (CAI)". CAI requires not only machine learning (of what exist) but machine creation (of what are new and useful). As we elaborated earlier, opportunity discovery and evaluation and design evaluation can be augmented by machine learning (e.g., supervised, or unsupervised models), and design generation can be augmented by machine creation (e.g., generative algorithms). Figure 2 summarizes the machine learning and machine creation methods to derive CAI in data-driven innovation processes. With CAI, even a novice engineer, designer or manager can work creatively on tasks that would otherwise require extensive knowledge acquisition, creative design skills, and gut feeling.

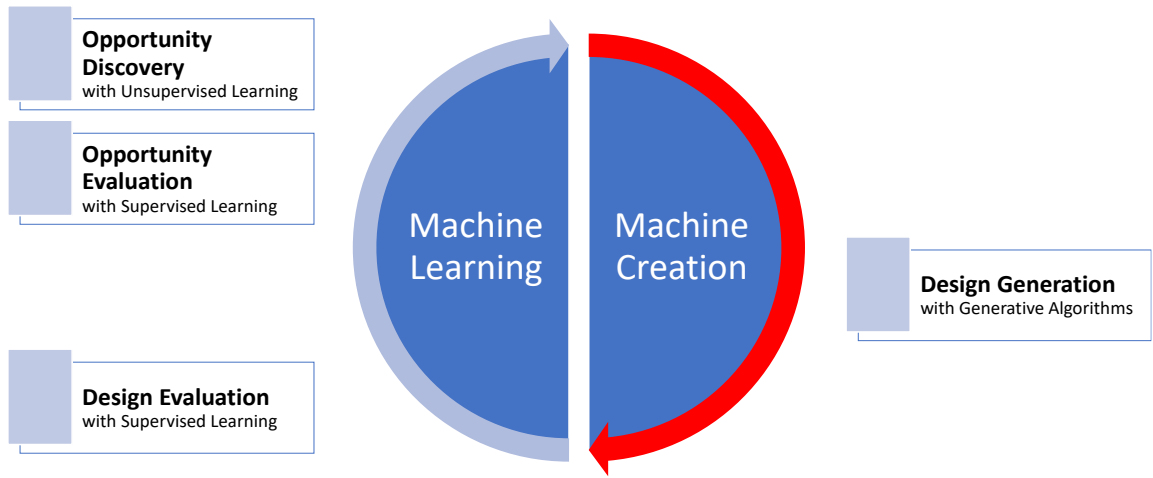

**Figure 2.** Creative AI requires both machine learning and machine creation

The future of innovation processes will be more data-driven and AI-empowered, while creative geniuses, formal R&D labs, open innovation activities and ecosystems will be continually important in the innovation process. Human innovators will play crucial roles as the developers and users of the data-driven innovation methods, tools, and expert systems, as well as the implementors of the resultant designs. In turn, these will make human innovators more informed and more inspired to discover more and better opportunities and generate more and better designs by themselves beyond what they could in purely intuitive or social processes. Figure 3 summarizes the roles of human innovators in the data-driven innovation process.

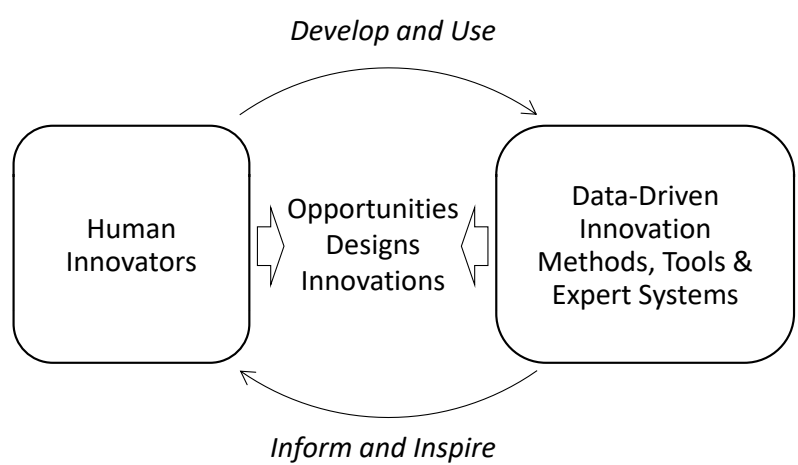

**Figure 3.** Roles of human innovators in the data-driven innovation process.

## 6. Embrace the Future toward Data-Driven Innovation

However, data-driven innovation is still at the early stage of development and adoption [11,18]. A search in Google for "data-driven innovation" today finds few true data-driven innovation research and practice as we have defined here,[1] but many in fact related to data-based innovation or data-driven optimization [1,2]. While companies are actively applying machine learning and AI to make their operations, products, and services more data-driven [6], few have pursued nor succeeded in data-driven innovation. Industrial firms tend to adopt data-driven decision making mainly for enhancing productivity [7], instead of creativity. There exist several impeding factors for the adoption of data-driven innovation by innovators and in organizations.

One bottleneck is the lack of readily available data-driven innovation methods, tools, and expert systems. Creative AI is under-developed in today's artificial intelligence R&D landscape, which is primarily focused on learning, classification, and pattern recognition tasks instead of creative ones. For the time being, companies and innovators may start with the existing machine learning and AI algorithms and customize them to support creative tasks in the innovation process. Moving forward, future research needs to draw on fundamental theories of creativity, design and innovation to develop new machine learning and machine creation algorithms specifically for innovation and to design integrative data-driven innovation methodologies, tools, expert systems, and workflows.

DDI requires data. To explore opportunities and generate new design concepts, innovators can creatively mine the public and free data sources, such as e-commerce sites, social media, open-source repositories (e.g., Github), patent and paper publication databases, and news archives. They can also mine and make use of those passively generated digital footprint data in their proprietary innovation activities, such as the records of prior failed experiments [33] or submissions or entries to their open innovation contests [34]. There are unbounded public or passively generated digital footprint data sources for innovators to mine, analyze and makes sense of for innovation. Despite additional investment required, companies may also redesign their workflows to purposefully generate and collect proprietary data streams to fuel their unique data-driven innovation needs.

---

[1] A Google search using the query of "Data-Driven Innovation" finds only two organizations focused on data-driven innovation research around the world. The first is the Data-Driven Innovation Lab at the Singapore University of Technology and Design, established in 2013. The other is the Data-Driven Innovation Initiative at the University of Edinburgh and Heriot-Watt University, established in 2018.

DDI relies on people and organizations that can seamlessly integrate domain expertise, innovation process knowledge, and mastery of data science and AI. Traditionally trained innovators might not be capable of or comfortable with using data science and AI to intelligentize the innovation process. The data scientists trained today are busy with applying their expertise to DDO and DBI, i.e., automating operations or increasing utility of user products and services, rather than the innovation process. For the long run, companies should collaborate with universities to educate and nurture future innovators with data science and AI knowledge and skills and future data scientists and AI experts who aspire to make the innovation process more data-driven and intelligent. In the short term, companies can assemble multidisciplinary DDI teams or retrain current specialists.

A strong leadership and organizational culture that embrace data-driven innovation beyond the traditional innovation processes are needed to support new data-driven innovation initiatives. There exist many more social-technical factors [35] that may enable or condition the development and adoption of data-driven innovation by individual innovators or organizations. Also, the data-driven innovation processes may take different forms in different organizations and networks of innovation agents [5]. Systematic research into the factors for the adoption of DDI and the structures and modes of human-AI interactions [36] in the innovation process will be desired to guide our journey toward the more data-driven and AI-empowered future of innovation processes.

The companies and organizations that start to experiment their own data-driven innovation processes and build relevant technological and organizational capabilities today may gain innovation advantages in the future. Meanwhile, the governments should also play a proactive role to facilitate the process, such as incentivizing or funding DDI-related talent and technological development, building infrastructure to enable public sector data accessibility to innovators and inter-organization data sharing. At the same time, policy makers should start to examine issues such as data privacy, security, ownership, civics, as well as the far-reaching DDI-induced economic, societal, and moral consequences and design policies and governance mechanisms to maximize its benefits and mitigate potential risks [11,37,38].

## 7. Concluding remarks

This article has made three major contributions. The first is the formal definition of data-driven innovation as an innovation process paradigm, distinguishing it from other data-related paradigms such as data-driven optimization and data-based innovation. The second is the taxonomy, based on the *double hump model*, to synthesize different data-driven innovation approaches. On this basis, several strategic recommendations are made for future research, education, practice, and governance to foster data-driven innovation. The definition and taxonomy of data-driven innovation as well as the strategic recommendations may guide future investments, capability building and policy development efforts toward data-driven innovation.